\documentclass{nature}
\usepackage{graphicx}

\title{The brightening of the pulsar wind nebula of PSR B0540--69 after its spin-down rate transition}

\author{M. Y. Ge$^{1}$, F. J. Lu$^{1}$, L. L. Yan$^{2}$, S. S. Weng$^{3}$, S. N. Zhang$^{1,4}$,
Q. D. Wang$^{5}$, L. J. Wang$^{1}$, Z. J. Li$^{6}$,  W. Zhang$^{1,4}$}

\begin{document}

\maketitle
\begin{affiliations}
\item Key Laboratory of Particle Astrophysics, Institute of High Energy Physics, Chinese Academy of Sciences, Beijing 100049, China
\item School of Mathematics and Physics, Anhui Jianzhu University, Hefei 230601, China
\item Department of Physics and Institute of Theoretical Physics, Nanjing Normal University, Nanjing 210023, China
\item University of Chinese Academy of Sciences, Chinese Academy of Sciences, Beijing 100049, China
\item Astronomy Department, University of Massachusetts, Amherst, MA 01003
\item School of of Architecture and Art, Hebei University of Engineering, Handan 056038, China
\end{affiliations}
\begin{abstract}

It is believed that an isolated pulsar loses its rotational energy mainly through a relativistic
wind consisting of electrons, positrons and possibly Poynting flux\cite{Pacini1973,Rees1974,Kennel1984}.
As it expands, this wind may eventually be terminated by a shock, where particles can be accelerated
to energies of X-ray synchrotron emission, and a pulsar wind nebula (PWN) is usually detectable
surrounding a young energetic pulsar\cite{Pacini1973,Rees1974,Kennel1984}. However,  the nature and/or
energetics of these physical processes remain very uncertain, largely because they typically cannot
be studied in a time-resolved fashion. Here we show that the X-ray PWN around the young pulsar
PSR B0540--69 brightens gradually up to 32$\pm8\%$ over the mean previous flux, after a sudden
spin-down rate ($\dot{\nu}$) transition (SRT) by $\sim36\%$\ in December 2011, which has very
different properties from a traditional pulsar glitch\cite{Marshall2015}. No evidence is seen for
any change in the pulsed X-ray emission. We conclude that the SRT results from a sudden
change in the pulsar magnetosphere that increases the pulsar wind power and hence the PWN
X-ray emission. The X-ray light curve of the PWN suggests a mean life time of the particles of
$397\pm374$\,days, corresponding to a magnetic field strength of
$0.78_{-0.28}^{+4.50}$\,mG in the PWN.

\end{abstract}

PSR B0540--69 is in the Large Magellanic Cloud, has a spin period of
50 ms, characteristic age of 1670\,yr, spin-down power $1.5\times10^{38}\,{\rm erg\,s^{-1}}$,
and is surrounded by a PWN\cite{Seward1984,Gotthelf2000,Petre2007}.
Those properties make PSR B0540--69 a twin of the famous Crab pulsar\cite{Petre2007}.
In the SRT in Dec. 2011, the spin-down rate (also called spin frequency derivative) $\dot\nu$ of PSR B0540--69  changed suddenly from
$-1.86\times10^{-10}{\rm\,Hz\,s^{-1}}$ to $-2.52\times10^{-10}{\rm\,Hz\,s^{-1}}$,
and has remained almost constant since then\cite{Zhang2001,Ge2012,Ferdman2015,Marshall2016}.
The pulsed X-ray flux and profile shape have remained unchanged since  the SRT, and no gamma-ray
variability with an upper limit of 30\% has been detected\cite{Marshall2015}.

\begin{figure*}
\begin{center}
\caption{The unfolded X-ray spectra of PSR B0540--69 and its
wind nebula (PSR+PWN) observed
by {\sl XMM-Newton}, {\sl Swift/XRT} and {\sl NuSTAR}. Panel (a): the red
points represent the unfolded spectrum taken by {\sl XMM-Newton} in 2000,
while the blue and green points correspond to
the unfolded spectra in 2015, by {\sl Swift/BAT} and {\sl NuSTAR}
respectively. The solid lines represent the best fitted power-law models.
To suppress the fluctuations, the data points for {\sl Swift/XRT} are merged from observations
in MJD 57070-57463 with a total exposure of 46.3\,ks. The {\sl NuSTAR} spectrum was taken
by two focal plane modules (usually labeled by FPMA  and FPMB) with an exposure of 59\,ks.
Panel (b): The residuals for spectral fitting. The uncertainties are at 1$\sigma$ level for all data points.}
\includegraphics[width=0.75\textwidth,clip]{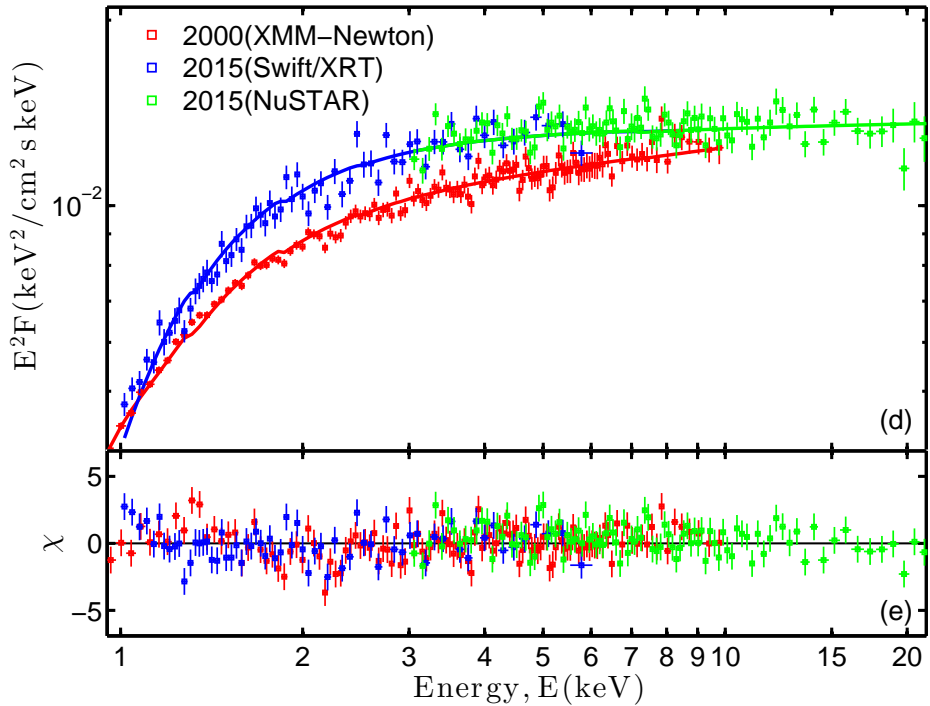}
\label{fig0}
\end{center}
\end{figure*}

\begin{figure*}
\begin{center}
\caption{The luminosity evolution of PSR B0540--69 and its wind nebula (PSR+PWN).
Panel (a): The data points represent the total X-ray luminosity of PSR+PWN
measured by {\sl XMM-Newton}, {\sl NuSTAR} and {\sl Swift/XRT}
in 1--10\,keV using a distance of 51\,kpc\cite{Gotthelf2000}.
The black dashed line is the spin-down luminosity of the central pulsar
multiplied by $7.09\%$, the ratio between the X-ray luminosity and
the spin-down luminosity before SRT, so as to compare them on the same scale. The blue solid line
is the fitted luminosity curve with equation (\ref{eq0}) and
$t_{d}=143\pm36$\,days\cite{Petre2007}. The vertical dotted line marks the SRT epoch\cite{Marshall2015}.
Panel (b): The data points are the pulsed luminosities of PSR B0540--69
in 1--10\,keV measured at several epochs, and the black dot-dashed line is the mean luminosity.
Panel (c): The photon indices of PSR+PWN obtained with different
instruments. The uncertainties are at 1$\sigma$ level for all data points.}
\includegraphics[width=0.9\textwidth,clip]{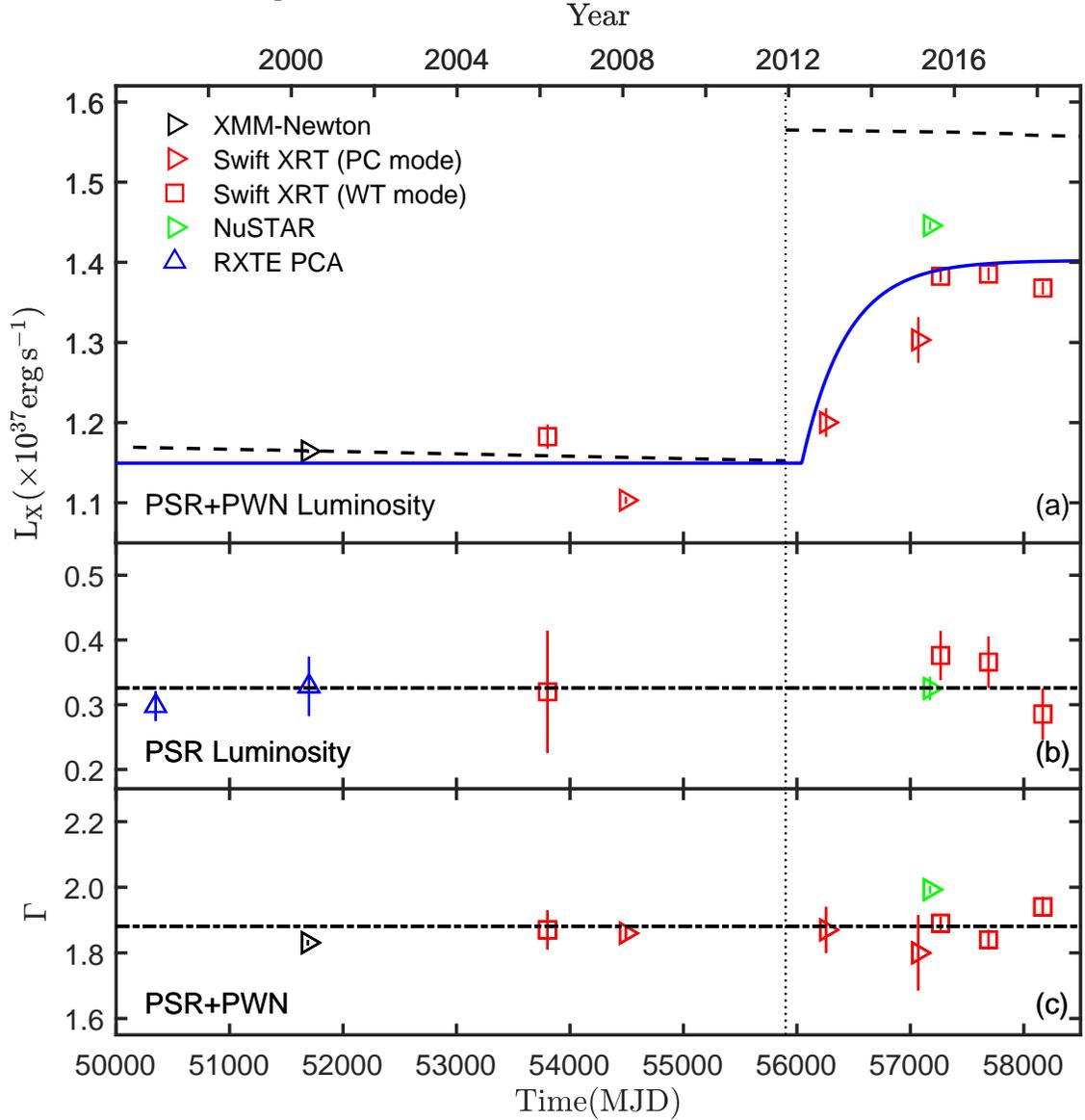}
\label{fig1}
\end{center}
\end{figure*}

We use observations from {\sl XMM-Newton}, {\sl NuSTAR} and the {\sl X-ray Telescope}
onboard the {\sl Neil Gehrels Swift Observatory (Swift/XRT)},
spanning from 1999 to 2018, to measure the total
X-ray luminosity of PSR B0540-69 plus the PWN (PSR+PWN) at
several epochs; these were measured together
because the two components cannot be resolved due to the limited angular
resolutions of these telescopes (refer to the ``Methods'' for the details of
analysis). As shown in Figures \ref{fig0}, \ref{fig1}
and Table \ref{table:3}, the spectra of PSR+PWN can be described by a power-law model
with flux increasing with time and a nearly constant spectral index.
The 1-10\,keV X-ray luminosity of the PSR+PWN has
 gradually increased  by $22\pm4\%$ ($1\sigma$, for this and all the
errors in this paper) over the mean pre-SRT value since December 2011.

The evolution of the total X-ray luminosity $L_{\rm X}$  post-SRT can be
characterized with a simple model (details in Methods):
\begin{equation}
L_{\rm X}=L_{\rm X0} [1 + \epsilon(1-e^{-(t-t_{1})/\tau})H(t-t_{1})],
\label{eq0}
\end{equation}
where $L_{\rm X0}$ is the X-ray luminosity in the SRT epoch $t_0$,
$\epsilon$ the ratio of the increased luminosity compared with
the luminosity at $t_{0}$, $\tau$ the variation time scale,
$t_{1}$ the time the PWN starts to brighten, which is $143\pm36$\,days after $t_0$
as inferred from the radius of the termination shock\cite{Petre2007}, and $H(t-t_1)$
a step function that equals to 0 when $t<{t_1}$, 0.5 when $t=t_{1}$  and 1 when $t>{t_1}$.
These parameters are obtained by fitting the X-ray luminosities on different dates, and the
results of which are listed in Supplementary Table 1. The best-fit value of
$\tau$ is $397\pm374$\,days.

Here we discuss the possible origins of the luminosity increase of the PSR+PWN after SRT.
We first investigate whether the pulsed flux of PSR B0540--69 varied from Feburary 2014 to September 2018,
and as shown in Figure \ref{fig1}(b), the pulsed flux remains almost constant, consistent
with the behaviors of this pulsar in the first month after the SRT\cite{Marshall2015}.
Secondly, from the spectral fitting results we know that ${\rm N_{H}}$, the absorption column density,
did not change  as a result of the SRT. This could not have induced the observed luminosity change.
Thirdly, we also found that the luminosity increase is unlikely due
to the brightening of recently heated ejecta. Observation of the {\sl High Resolution Camera}
onboard the {\sl Chandra X-ray Observatory} revealed that the X-ray flux of the supernova
remnant surrounding PSR B0540-69 is about 18\% of that of the PSR+PWN\cite{Gotthelf2000}.
If the luminosity increase were due to the recently heated ejecta, the X-ray flux of the supernova
remnant should have doubled in several years, which is almost impossible given the fact that
the system has a characteristic age of 1670 years.
We therefore conclude that the only possible origin of the observed luminosity increase
of the PSR+PWN is due to the brightening of the PWN. From the combined PSR+PWN
flux and the pulsed flux (refer to Figure \ref{fig1} and  Table \ref{table:4}), it can be
inferred that the 1-10\,keV X-ray luminosity of the PWN has increased by $32\pm8\%$.

We note that the SRT is unlike traditional pulsar glitches. The latter refer to
sudden pulse frequency and/or frequency derivative changes
generally interpreted as the result of angular momentum
transfer between the inner superfluid material and the outer crust\cite{Alpar1984a}.
In the SRT, the spin frequency ($\nu$) shows no sudden change, while $\dot\nu$ changes by 36\% within 2 weeks
and has remained nearly constant over the past 6.5 years (refer to {\bf Supplementary Figure 1 and Supplementary Table 2}).
These are obviously different from the traditional glitches, which, in addition to the sudden
pulse frequency, and often frequency derivative changes, also show tendency of
recovering to pre-transition values on a time scale of $\sim$ 100\,days\cite{Espinoza2011}.
The evolution of this pulsar's spin after
the SRT is instead similar to the long-term evolution of a typical pulsar, and the accelerated
spin-down since the SRT is thus experienced by the pulsar as a whole, rather than solely by the crust.
Therefore, the pulsar's spin-down power $L_{\rm sd}$ can be expressed by
the following canonical equation:
\begin{equation}
L_{\rm sd}=4\pi^{2}I\nu\dot{\nu},
\label{eq1}
\end{equation}
where $I$ is the moment of inertia of the pulsar.
According to equation (\ref{eq1}) and the values of $\nu$ and $\dot{\nu}$ pre- and post-SRT,
$L_{\rm sd}$ of PSR B0540--69 should increase by
$36\%$ after SRT.  This is near to the fractional increase of the
PWN flux and implies that the brightness of the PWN is directly related to $L_{\rm sd}$.
The physical connection between $L_{\rm sd}$ increase and PWN brightening should be the pulsar wind strength.

In the widely accepted pulsar wind and PWN scenario, the pulsar wind
leaves the pulsar at almost the speed of light, but only
starts to emit after
being randomized and probably further accelerated at the termination
shock\cite{Pacini1973,Rees1974,Kennel1984}.
We denote the time taken for the particles moving from
the pulsar to the termination shock as $t_{\mathrm{TS}} \simeq R_{\mathrm{TS}}/c$,
where $R_{\mathrm{TS}}=0.12\pm0.3$\,pc is the distance of the termination shock from the pulsar\cite{Petre2007}.
The pulsar wind strength increases immediately after the SRT, but the PWN
starts to brighten at $t_{\mathrm{TS}}=t_d=t_{1}-t_{0}\sim143\pm36$\,days. Then,
given the synchrotron emission nature of the
PWN\cite{Kennel1984,Bucciantini11,Kargaltsev15,Wang16a},
we can estimate the magnetic field of the
PWN by equating the synchrotron time scale $\tau_{\rm s}$ to the rising time $\tau$ of the PWN emission.
Using $\tau_{\rm s}\sim446.6B_{\rm mG}^{-3/2}E_{\rm keV}^{-1/2}$\,days,
where the magnetic field strength $B$ is in milli-Gauss (mG) and the photon energy in keV\cite{Petre2007},
we can infer that the magnetic field strength is  $0.78_{-0.28}^{+4.50}$\,mG, where a photon
energy of 2.66\,keV (the weighted mean of the X-ray photons) is used.
This inferred magnetic field strength is consistent with the equipartition
value, $\sim$0.8\,mG of this PWN\cite{Petre2007}.

The brightening of the PWN of PSR B0540--69 is unique, though
flux variations have also been detected in other PWNe. Among these PWNe,
one of the most extensively studied cases is the Crab nebula, which
has GeV flares and
whose X-ray flux  changes by up to 10\%  over a year-timescale\cite{Tavani2011,Abdo2011}.
The PWN associated with PSR J1846--0258 is expanding but has dimmed by $\sim10\%$ between
2000 and 2016 (see ref. \citen{Reynolds2018}).  However, neither the Crab pulsar nor
PSR J1846-0258 have shown corresponding $L_{\rm sd}$ changes.
The variations of the Crab nebula are interpreted as being due to
the instability of the PWN rather than
induced by the central pulsar\cite{Kargaltsev15}. The fading of the PWN of
PSR J1846--0258 has not been well explained, but unlikely connected with
changes of $L_{\rm sd}$, which has been nearly steady throughout\cite{Reynolds2018}.
Other PWNe, such as that of the Vela pulsar, show local X-ray instability\cite{Pavlov2003}
{, but again}, no connection between the instability
and the rotation state changes of the central pulsars has been found.
Hence, the PWN brightening of PSR B0540--69 appears to be very different
from these phenomena, as it seems to have been triggered by the variation
of the spin-down state of the central pulsar.

Comparison between the SRT behaviors of PSR B0540--69 and the timing
variabilities of other pulsars may provide us with information on the
mechanism producing the SRT. Lyne et al. 2010 found that in six pulsars the
changes in $\dot{\nu}$ ascribed to timing noise are correlated with the
variations of their radio pulse profiles over timescales
of about one to several years\cite{Lyne2010}. They thus proposed that the
timing instability is the result of changes in the pulsar magnetosphere.
Among these pulsars, PSR B2035+36 has the most similar timing behaviors to the
SRT of PSR B0540--69 with respect to the amplitude (13.28\%) of $\dot{\nu}$
variation and the time ($>1800$\,days) over which it remains in the large $\dot{\nu}$ state.
Besides these six pulsars, PSR B1931+24 and PSR J2021+4026 also have
similar timing behaviors with the SRT of PSR B0540--69. PSR B1931+24 has a $\dot{\nu}$ variation
amplitude of  44.90\%, but it is an intermittent pulsar with
dramatic flux changes\cite{Kramer2006}.
In October 2011, PSR J2021+4026 experienced a sudden change {\bf in} $\dot{\nu}$ by 4\%
but no significant change in $\nu$, and at the same time
significant gamma-ray flux and profile variations were detected\cite{Allafort13},
which are attributed to changes in the magnetosphere\cite{Allafort13}. The SRT
of PSR B0540--69 is similar to the timing  instabilities of these pulsars, because
they all show fast state  transitions and appear to remain in the new states
for months to years. Given these similar timing behaviors and the different
characteristics to a traditional pulsar glitch, where the latter is ascribed to changes
in the pulsar crust and/or interior, we believe that the SRT
of PSR B0540--69 is also due to the variations of its magnetosphere.
However, the exact changes to the magnetosphere of PSR B0540--69
and the observational consequences are probably different from the other
pulsars: its X-ray emission regions have not been affected significantly,
so pulse profile or pulsed flux variations {\bf seen} in the other pulsars are
not detected in this pulsar (refer to Figure 2 and Supplementary Figure 2);
instead, the changes to the magnetosphere of
PSR B0540-69 result in the brightening of the PWN in the X-ray band.

According to the classic structure of a pulsar, the pulsed emission is
produced in
the co-rotating magnetosphere and the pulsar wind is produced outside the
light cylinder, which is connected with the pulsar through open magnetic
field lines originating from the magnetic polar region\cite{Goldreich1969}. Theoretical studies
show that the radio and X-ray emissions of a pulsar probably come from
different regions:  the radio-emitting region is closer to the magnetic polar
region\cite{Ruderman1975,Arons1979}, but the X-ray emission is
from a broader region such as the outer gap\cite{Cheng1986}. We thus
suggest that the major changes to the magnetosphere of PSR B0540-69 take
place in the magnetic polar region. If PSR B0540-69 has pulsed radio emission,
it may also have detectable profile vairations following the SRT, similar to those
observed by Lyne et al. 2010 (see ref. \citen{Lyne2010}).

The varying braking index and the unchanged pulsed emission of
PSR B0540-69 after the SRT
set important constraints on theoretical studies of pulsar magnetospheres.
The braking index of PSR B0540--69 changed from ${\rm n}=2.129\pm 0.012$ pre-SRT
to ${\rm n}=0.031\pm 0.013$ in the first four years after the SRT\cite{Marshall2016},
and then to its current value ${\rm n}=0.94\pm 0.01$ (refer to the Supplementary Figure 1 and Supplementary Table 2).
Kou et al. 2016 proposed that the increased pulsar wind strength could lead to a small
braking index\cite{Kou2016}. However, their model cannot give a quantitatively satisfactory
explanation to the post-SRT braking index and the enhancement of the pulsar wind.
Ek\c{s}i suggested that the low braking index of PSR B0540-69 post-SRT is due to the diffusion
of the magnetic fields buried in the crust by rapid fallback accretion soon after the neutron star
birth\cite{Eksi17}. If such a model is correct, the magnetic field is probably buried in the pulsar magnetic
pole region so that the increased magnetic field would mainly enhance the wind production outside
the light cylinder. It has also been realized that there are
current sheets separating the open from the closed regions of the magnetosphere\cite{Bucciantini06}.
In this scenario, the braking index is\cite{Bucciantini06}
\begin{equation}
{\rm n}=3+2\frac{\partial \ln \left( 1+\frac{R_{L}}{R_{Y}}\right) }{\partial \ln
\Omega },
\label{eq2}
\end{equation}
where $R_L$ is the radius of the light cylinder, $R_Y$ the size of the
closed regions of the magnetosphere. To yield a braking index ${\rm n}\sim0$, $R_Y$ should
change significantly, which seems to conflict with the undetectable changes of the pulsed X-ray emission.
However, equation (\ref{eq2}) predicts the consequence of overall magnetosphere
variations, and as discussed above, the changes in the magnetosphere of PSR B0540--69
may mainly be in the magnetic polar region, which has no major influence on the pulsed X-ray emission of PSR B0540--69.

With the direct evidence that pulsar wind is currently the main force
braking the spin of this pulsar, more theoretical work is needed to
simultaneously explain the strengthening of the wind, the lack of
changes to the pulsed X-ray profile and flux, as well
as the $\sim0$ braking index of the pulsar post-SRT. New X-ray
observation with high spacial resolution is also urged to test the scenario
for the SRT and to explore the dynamics of the pulsar outflows, by
comparing the new observation to the archive data obtained by the {\sl Chandra
X-ray Observatory} before the SRT.


\clearpage

\clearpage

\begin{table}
\footnotesize
\caption {Spectral fitting results of PSR B0540--69 and its wind nebula
(PSR+PWN) with observations of {\sl XMM-Newton}, {\sl Swift/XRT} and {\sl NuSTAR}. The
uncertainties are at 1$\sigma$ level for all the parameters.}
\scriptsize{}
\label{table:3}
\medskip
\begin{center}
\begin{tabular}{c c c c c c c}
\hline \hline
Telescope                             &Time & ${\rm N_{H}}$   & Photon Index  & Flux (1-10 keV)     &reduced $\chi^2$ (d.o.f.)\\
                                            & MJD  & 10$^{21}$\,cm$^{-2}$ &            &  10$^{-11}$\,erg\,s$^{-1}$\,cm$^{-2}$&\\
\hline
 {\sl XMM-Newton}              &  51690   & $3.4\pm0.1$ &	$1.83\pm0.01$	&	$3.96\pm0.01$ &	 1.12(152)  \\
\hline
 {\sl NuSTAR}                  & 57173    & 3.4        & $1.99\pm0.01$     &   $4.91\pm0.04$ & 1.08(197) \\
\hline
  {\sl Swift/XRT} (PC mode) & 53474-55511 &	$4.1\pm0.1$   &	$1.86\pm0.01$   &   $3.75\pm0.03$ &	 0.98(2871) \\
  {\sl Swift/XRT} (PC mode) & 56254-56261 &	$4.7\pm0.6$   &	$1.87\pm0.04$	&   $4.08\pm0.10$ & 1.26(193) \\
  {\sl Swift/XRT} (PC mode) & 56967-57173 &	$4.1\pm1.0$   &	$1.80\pm0.05$	&   $4.43\pm0.16$ & 1.16(88) \\
  {\sl Swift/XRT} (WT mode) & 53761-53843 &	$3.5\pm0.5$   &	$1.87\pm0.03$   &	$4.02\pm0.09$ &1.05(259) \\
  {\sl Swift/XRT} (WT mode) & 57070-57463 &	$3.7\pm0.2$   &	$1.89\pm0.01$   &	$4.70\pm0.04$ &1.05(1533) \\	
  {\sl Swift/XRT} (WT mode) & 57493-57884 &	$3.5\pm0.2$   &	$1.84\pm0.01$   &	$4.71\pm0.05$ &1.01(1396) \\
  {\sl Swift/XRT} (WT mode) & 57915-58420 &	$3.9\pm0.2$   &	$1.94\pm0.02$	&   $4.65\pm0.05$ &	 1.04(952) \\
\hline
\hline \hline
\end{tabular}
\end{center}
\end{table}

\begin{table}
\footnotesize
\caption {Joint spectral fitting results of PSR B0540--69 for observations by
{\sl RXTE/PCA}, {\sl Swift/XRT} and {\sl NuSTAR}( refer to Methods for spectral model information). The
uncertainties are at 1$\sigma$ level for all the parameters.}
\scriptsize{}
\label{table:4}
\medskip
\begin{center}
\begin{tabular}{c c c c c c c}
\hline \hline
Telescope    & Time      & ${\rm N_{H}}$  & $\alpha$ &$\beta$  & Flux (1--10\,keV)    &reduced $\chi^2$ (d.o.f.) \\
                     & MJD      & 10$^{21}$\,cm$^{-2}$ &  &   &  10$^{-11}$\,erg\,s$^{-1}$\,cm$^{-2}$&\\
\hline
{\sl RXTE/PCA}    & 50348    & 3.4  & $1.11\pm0.07$  &$0.56\pm0.06$ &   $1.02\pm0.08$ & 1.11(1725)  \\
                             &      51700       & --  & --  & -- & $1.13\pm0.16$ & -- \\
\hline
{\sl NuSTAR}        &    57173        & --  & --   &--& $1.11\pm0.05$ & -- \\
\hline
{\sl Swift/XRT} (WT mode) &  53761-53843 & --  & -- &--&   $1.09\pm0.33$ &  -- \\
{\sl Swift/XRT} (WT mode) &  57070-57463 & --  & --  &--&   $1.27\pm0.12$ & -- \\
{\sl Swift/XRT} (WT mode) &  57493-57884 & --  & --  &--&   $1.24\pm0.13$ & -- \\
{\sl Swift/XRT} (WT mode) &  57915-58420 & --  & --  &--&   $0.97\pm0.13$ & -- \\
\hline \hline
\end{tabular}
\end{center}
\end{table}

\clearpage

{\bf Methods}

{\bf Data Reduction and Analyses}

We select the X-ray telescopes {\sl XMM-Newton}, {\sl NuSTAR} and {\sl Swift/XRT} to
measure the flux of PSR B0540--69 and its wind nebula (PSR+PWN).
Due to the relatively large distance, the central pulsar and its nebula
can not be individually resolved spatially, hence only the overall fluxes
of the PSR+PWN are obtained by these observations. In order to estimate
the pulsed X-ray flux of PSR B0540--69, we only utilize the observations with good time resolution,
such as by {\sl RXTE/PCA}, {\sl NuSTAR}, and {\sl Swfit/XRT} in
the Windowed Timing (WT) mode. We use the pulse-on spectra with the background from the
pulse-off region subtracted to estimate the pulsed flux
of the central pulsar\cite{Ge2012}. Each spectrum is fitted
by an absorbed power-law model ({\it tbabs*powerlaw} in XSPEC).
Due to the relatively low count rate of the pulsar, in the spectral analyses
of the pulsed emission, the absorption column density ${\rm N_{H}}$
is fixed to 3.4$\times10^{21} {\rm cm}^{-2}$, the value obtained by {\sl XMM-Newton}
when fit the overall spectrum of the PSR+PWN.
The coordinates of PSR B0540--69 are
$\alpha=05^{\rm h}40^{\rm m}11^{\rm s}.204$ and
$\delta$=-69$^{\textordmasculine}$19$^{\prime}$54$^{\prime\prime}$.34
(J2000)\cite{Mignani2010}, which are used to correct the arrival
time of each photon to the Solar System Barycentre (SSB).

\subsection{{\sl RXTE/PCA}}
\subsection{}

The data collected by the Proportional Counter Array (PCA) onboard the
{\sl Rossi X-Ray Timing Explorer (RXTE)} were used in this paper.
PCA is one of the main instruments operating in 2 to 60\,keV
with time resolution up to 1 ${\rm \mu}$s\cite{Jahoda2006}.
Altogether there are 1448 observations on PSR B0540--69 with a total
exposure time of 3481\,ks, which make the detailed temporal studies
possible.

The Good Xenon mode data from observations P10206, P10218,
P10250, P20188, P30087, P40139, P50103, P60082, P70092, P80089,
P80118, P90075, P91060, P92010, P93023, P93448, P94023, P95023 and P96023
are selected for our timing analyses, and then the Standard {\sl RXTE} data
processing method is used to obtain the timing data as follows.
(1) Generate the Good Time Interval by the FTOOL command \texttt{maketime}
based on the {\sl RXTE} filter file; (2) merge the Good Xenon data with
\texttt{make\_se} and then filter the events with \texttt{grosstimefilt};
(3) convert the arrival time of each photon to the SSB with \texttt{faxbary}.
Among these observations, P10206 and P50103 were used to
create the pulsed spectra as they are the only two on-axis observations
with relatively long exposure time.

\subsection{{\sl XMM-Newton}}
\subsection{}

We processed the data collected by the European Photon Imaging Camera (EPIC)
of {\sl XMM-Newton} around May 26, 2000 (see ref. \citen{Jansen2001}, ObsID 0117510201, exposure 10\,ks)
using the Science Analysis System (\texttt{SAS}) (v14.0.01) software . The time
intervals contaminated by flaring particle background are discarded.
Events in a circular region with a radius of $50^{\prime\prime}$ centred
on the pulsar position are selected to ensure that all the source events
are included, while the background region is a co-centred annulus
with the inner and outer radii of $50^{\prime\prime}$ and $80^{\prime\prime}$, respectively. The
instrument response files for spectral analyses are generated
by the \texttt{SAS} commands \texttt{rmfgen} and \texttt{arfgen}, and
the spectra are grouped to have at least 15 counts per bin with the
command \texttt{specgroup}, to enable the use of chi-square statistics.
The region with radius less than $2.5^{\prime\prime}$ is subtracted
to preclude the pile-up effect, and the effect of this subtraction is
automatically corrected by the software ( http://xmm-tools.cosmos.esa.
int/external/xmm\_user\_support/documentation/uhb/epic\_pile-up.html).

\subsection{{\sl NuSTAR}}
\subsection{}

PSR B0540--69 was observed by {\sl NuSTAR}\cite{Harrison2013} around May 30
(OBSID 40101013002), 2015, with an exposure time of 59\,ks. In
this work, we analyze data from two telescopes on NuSTAR (usually labeled by their
focal plane modules, FPMA and FPMB) using \textsc{HEASoft} (version 6.24).
We utilize \texttt{nupipline} with {\sl NuSTAR} CALDB v20180312 to
create GTIs and select a circular region of radius $180^{\prime\prime}$ centred on the pulsar position to
extract the source spectrum. The background spectrum was extracted from a circular
region of radius 150$^{\prime\prime}$ centred on  $\alpha=05^{\rm h}38^{\rm m}54^{\rm s}.997$ and
$\delta$=-69$^{\textordmasculine}$19$^{\prime}$30$^{\prime\prime}$.683, which
is source free and not far away from PSR B0540--69.
The arrival time of each photon is corrected to SSB too.

\subsection{{\sl Swift/XRT}}
\subsection{}

The {\sl Neil Gehrels Swift Observatory (SWIFT)}\cite{Gehrels2004} is suitable for
monitoring the variability of X-ray sources owing to its rapid slew capability.
Until May 28, 2018, there are 109 observations pointing to the position of
PSR B0540--69, of which 77 were taken in the WT mode
(00033603, 00053402, 00093144) and 32 in the Photon Counting (PC)
 mode (00053400, 00053402, 00033490,00045451--3)\cite{Campana2008}.
 The exposure time of an individual
{\sl Swift} observation ranges from 254\,s to 20\,ks, with a mean value of 3.2\,ks. All
the data are analyzed with the packages in \textsc{Heasoft} version
6.24 ( https://heasarc.gsfc.nasa.gov/docs/software/lheasoft/).
The initial event cleaning is executed with the command \texttt{xrtpipeline}
using the standard quality cuts.

Because the count rate of PSR B0540--69 is $\sim0.75$\,cts\,s$^{-1}$,
the PC mode data suffer from pile-up problems. We
therefore extract the source spectra from a circular region with a radius of 20\,pixels (47.15$^{\prime\prime}$)
but exclude the central 2 pixels that have the highest count rates
to minimize the contamination of
pile-up (refer to the Supplementary information and
https://heasarc.gsfc.\\nasa.gov/docs/heasarc/caldb/swift/docs/xrt/XRT-OAB-CAL\_ARF\_v3.pdf).
The background \\ spectra are extracted from an annulus region with the inner and
outer radii of 20 and 40\,pixels, respectively. The ancillary response files (ARFs)
and the corresponding response files (RMFs) are generated separately.
ARFs are produced with the command \texttt{xrtmkarf}, while
the corresponding RMFs are taken from the CALDB database.
The spectra are regrouped to have at least 20 counts per bin
to ensure that the
$\chi^2$ statistics can be used. In order to provide better constraints on
the spectral parameters, we also divide the observations
into several groups and fit the spectra simultaneously for
each group.

Thanks to the good time resolution ($\sim1.8$\,ms), the WT mode data are
adequate for timing analysis. In the 77 WT mode observations, we exclude
the observation performed on March 04 2016, because the
source is erratically located in and out of
the field of view\cite{Campana2008}. The source events in 1--10\,keV are
extracted from a circular region with a radius of 15 pixels
centred on the source position, and the background spectrum is extracted from
an annulus with the inner and outer radii of 15 and 30 pixels, respectively.
The photon arrival time is converted to SSB using the
FTOOL command \texttt{barycorr}. Light curves with a bin size of 2\,ms are
extracted and corrected for the telescope vignetting and point spread function
effects by running the {\sl Swift} script \texttt{xrtlccorr}. The
background-subtracted light curves are plotted in Supplementary Figure 4.

As shown in Supplementary Figure 4, the {\sl Swift/XRT} count rate
of PSR+PWN increases gradually after SRT while the pulsed count rate
does not show any significant enhancement. Then, the count rate
variation of PSR+PWN is also fitted with equation (\ref{eq1}),
and the parameters are listed in Supplementary Table 2, which are
consistent with those obtained from the luminosities by different instruments.

{\bf Timing and pulsed emission }

The timing analysis of PSR B0540--69  pre-SRT has been performed
in details by Ferdman et al. 2015 (see ref. \citen{Ferdman2015}).
So we adopt the pre-SRT timing parameters from their work, and only
focus on the  post-SRT timing analyses.

We calculate the rotation frequencies and their derivatives from
the pulse times-of-arrivals (TOAs) of the pulsar in different epochs. To
obtain the TOAs, we first create an accurate template profile by adding
all the aligned profiles from our full data set in the entire energy band for
each instrument. The template contains
10\,bins because of the low count rate of PSR B0540--69.
We then obtain
the TOAs for each observation by comparing the template profile with the one
from that observation, as detailed in the following part.
(1) Search for the spin frequency that makes the folded profile deviating the
most from a uniform distribution as represented by the  Pearson $\chi^2$ value,
and this is taken as the best spin frequency\cite{Ge2012}. (2) Fold the pulse
profile with the starting time of the observation as the reference epoch of the TOA obtained in the next step.
(3) Calculate the phase shift using the cross-correlation
between the pulse profile and template profile, from which we can calculate the
TOA representing each observation in the reference epoch. The uncertainty of a TOA is
calculated with a Monte-Carlo method\cite{Ge2012} as follows.
(1) Generate a fake profile based on the Poisson count
error of each phase bin for the corresponding integrated profile;
(2) Calculate
the phase shift using the cross-correlation between the fake profile
and the template profile; (3) Repeat steps (1) and (2) for 400 times
and obtain 400 fake TOAs; (4) The 1$\sigma$ width of the distribution
of the 400 fake TOAs is taken as the uncertainty of the TOA in one observation.

We perform both part-timing analysis and fully coherent timing analysis for
PSR B0540--69 to study its timing behaviors using TEMPO2\cite{TEMPO2}, similar to Ferdman
et al. 2015 (see ref. \citen{Ferdman2015}). The part-timing analyses with a typical duration
of 100 days were performed for all the post-SRT observations, to obtain
the long-term evolution of the rotation frequency, while the fully coherent
timing analyses were performed only for three time intervals, MJD 57070--57546,
57546--57946 and 57946--58420 to obtain more accurate pulse profiles,
as two small glitches on MJD 57546 and 57946 are detected  at the
boundary epochs of the three time ranges\cite{Mignani2018}.
In equation (\ref{eq10})  below, which describes the pulse phase calculation,
we only consider $\nu$ and $\dot\nu$ in the part timing analysis
because of the relatively short duration, and include $\ddot\nu$ in
the fully
coherent timing analysis.
\begin{equation}
\Phi-\Phi_{0}=\nu{(t-t_{e})}+\frac{1}{2}\dot\nu{(t-t_{e})}^{2}+\frac{1}{6}\ddot\nu{(t-t_{e})}^{2}
\label{eq10}
\end{equation}
where $\Phi_{0}$, $\nu$, $\dot\nu$ and $\ddot\nu$ are the phase, frequency,
frequency derivative and the second order derivative of frequency at the
reference epoch
$t_{e}$, respectively. The resultant timing residuals are less than 1.4\,ms and
distributed quite smoothly as shown in Supplementary Figure 3 and Supplementary Table 2,
implying that the inferred
rotation parameters are reliable. It can be also seen that $\dot\nu$ only changes
very slightly, which means that $L_{\rm sd}$ keeps almost constant post-SRT.
However, with the new value of $\nu$ and $\ddot{\nu}$, it can be inferred
that the braking index increases from $<0.1$ in the first four years after
SRT\cite{Marshall2016} to $0.94\pm0.01$ in the recent two years.

With the timing parameters obtained above, we can determine the pulsed X-ray profiles and
the fluxes during different epochs. As shown in Supplementary Figure 2, the pulse profile of PSR B0540--69
remains almost unchanged post-SRT. To get the pulsed flux, the events
in phase range 0.4--0.6 are used to generate the background
spectra and those
outside this phase range are taken to generate the source spectra. Because of the low count
rate of the pulsar, the pulsed spectra are fitted with a {\it log-parabola} model\cite{Campana2008}
by freezing  ${\rm N_H}$ to 3.4$\times10^{21}\,{\rm cm^{-2}}$,
forcing all the photon indices equal to each other and only setting the normalization
values of different spectra as free parameters. The resultant spectral parameters and fluxes
are listed in Table \ref{table:4}. The averaged pulsed flux pre-SRT is
$(1.04\pm0.07)\times10^{-11}$\,ergs\,s$^{-1}$\,cm$^{-2}$,
while the averaged pulsed flux
is $(1.13\pm0.04)\times10^{-11}$\,ergs\,s$^{-1}$\,cm$^{-2}$
post-SRT. Therefore, the pulsed flux remains unchanged within about
1$\sigma$, consistent with the results of Marshall et al. 2015 (see ref. \citen{Marshall2015}).

{\bf Derivation of Equation (1)}

As shown by many studies, the pulsar wind is lepton-dominated
at the termination shock\cite{Kennel1984}.
For simplicity, we assume that the Lorentz factor (velocity) of
the electrons/positrons keeps unchanged post-SRT, because the spectral
indices pre- and post-SRT are the same. In this case the electron/positron
injection rate $I_{e}$ is proportional to the
spin-down luminosity of the pulsar
\begin{equation}
I_{e}=fL_{\mathrm{sd}},
\end{equation}
where $f$ is the converting factor. Denoting the total number of electrons/positrons in the
PWN as $N_{e}$, we have
\begin{equation}
I_{e}-\frac{N_{e}}{\tau}=\frac{dN_{e}}{dt},  \label{eq:Ne-evolve}
\end{equation}
where the second term on the left side is the electron/positron number dying
away because of synchrotron emission and $\tau$ is the mean lifetime of the
electrons/positrons.

Further assuming that the magnetic field in the PWN keeps constant
during the SRT, the X-ray luminosity of the PWN is proportional to the total
number of electrons/positrons, which is
\begin{equation}
N_{e}=\alpha L_{X},
\end{equation}
by which Equation $\left( \ref{eq:Ne-evolve}\right) $ can be rewritten as
\begin{equation}
\frac{fL_{\mathrm{sd}}}{\alpha }-\frac{L_{X}}{\tau}=\frac{dL_{X}}{dt}.
\end{equation}
As $L_{\mathrm{sd}}$, $\tau$, $f$ and $\alpha$ are all constants,
the above equation can be readily integrated to give the
desired result
\begin{equation}
L_{X}=L_{X0}+\left( L_{X,\mathrm{new}}-L_{X0}\right) \left[ 1-e^{-\left(
t-t_{1}\right) /\tau}\right] ,
\end{equation}
where $L_{X,\mathrm{new}}=\tau fL_{\mathrm{sd}}/\alpha $ is the saturated
X-ray luminosity of the PWN after the SRT, and $t_{1}$ is the time at which
the X-ray luminosity starts to increase.
The original X-ray luminosity of the PWN is $L_{X0}=\tau fL_{\mathrm{
sd,0}}/\alpha $, where $L_{\mathrm{sd,0}}$ is the spin-down luminosity of
the pulsar before the SRT. We could therefore re-write the formula as,
\begin{equation}
L_{X}=L_{X0}[1+{\epsilon}(1-e^{-(t-t_{1})/\tau})],
\end{equation}
where ${\epsilon}=\left(L_{X,\mathrm{new}}-L_{X0}\right)/L_{X0}$.

{\bf Data availability}

The data that support the plots within this paper and other findings of this study
are available from the corresponding author M.-Y.G.(gemy@ihep.ac.cn) upon
reasonable request. All the observational data used in this study are public and
can be downloaded from the archives of these X-ray satellites.

{\bf Acknowledgements} We thank the anonymous referees for their insightful suggestions.
Professors K. S. Cheng of the Hongkong University and L. Zhang of the Yunnan University
are appreciated for helpful discussions on the emission mechanism of pulsars. This work
is supported by the National Key
R\&D Program of China (2016YFA0400800)  and the National Natural
Science Foundation of China under grants 11503027,
 11673013, 11653004, U1838201, U1838202 and U1838104.
We thank the data support from {\sl XMM-Newton}, {\sl NuSTAR}, {\sl RXTE} and {\sl Swift} teams.

{\bf Author contributions}
M.-Y.G., L.-L.Y., S.-S.W., Z.-J.L. and W.Z. were involved in the data analysis.
F.-J.L., M.-Y.G., L.-J.W., S.-N.Z., and Q.-D.W. contributed to the theoretical discussions.
The manuscript was produced by M.-Y.G., F.-J.L., L.-J.W., Q.-D.W., S.-N.Z. and S.-S.W..

{\bf Competing Interests} The authors declare that they have no competing financial interests.

{\bf Correspondence} Correspondence and requests for materials should be addressed to (F.J.L and M.Y.G, email:  lufj@ihep.ac.cn, gemy@ihep.ac.cn).


\clearpage


{\bf References}
\begin{thebibliography}{10}

\expandafter\ifx\csname url\endcsname\relax
  \def\url#1{\texttt{#1}}\fi
\expandafter\ifx\csname urlprefix\endcsname\relax\def\urlprefix{URL }\fi
\providecommand{\bibinfo}[2]{#2}
\providecommand{\eprint}[2][]{\url{#2}}

\bibitem{Pacini1973}
\bibinfo{author}{{Pacini}, F. and {Salvati}, M.}
\newblock \bibinfo{title}{On the Evolution of Supernova Remnants. Evolution of the Magnetic Field, Particles, Content, and Luminosity}
\newblock \emph{\bibinfo{journal}{Astrophys. J.}}
  \textbf{\bibinfo{volume}{186}}, \bibinfo{pages}{249-266}
  (\bibinfo{year}{1973}).

\bibitem{Rees1974}
\bibinfo{author}{{Rees}, M.~J. and {Gunn}, J.~E.}
\newblock \bibinfo{title}{The origin of the magnetic field and relativistic particles in the Crab Nebula}
\newblock \emph{\bibinfo{journal}{Mon. Not. R. Astron. Soc.}}
  \textbf{\bibinfo{volume}{167}}, \bibinfo{pages}{1-12}
  (\bibinfo{year}{1974}).


\bibitem{Kennel1984}
\bibinfo{author}{{Kennel}, C.~F. and {Coroniti}, F.~V.}
\newblock \bibinfo{title}{Magnetohydrodynamic model of Crab nebula radiation}
\newblock \emph{\bibinfo{journal}{Astrophys. J.}}
  \textbf{\bibinfo{volume}{283}}, \bibinfo{pages}{710-730}
  (\bibinfo{year}{1984}).

\bibitem{Marshall2015}
\bibinfo{author}{{Marshall}, F.~E. and {Guillemot}, L. and {Harding}, A.~K. and {Martin}, P. and {Smith}, D.~A.}
\newblock \bibinfo{title}{Discovery of a Spin-down State Change in the LMC Pulsar B0540-69}.
\newblock \emph{\bibinfo{journal}{Astrophys. J. Lett.}}
  \textbf{\bibinfo{volume}{807}}, \bibinfo{pages}{L27}
  (\bibinfo{year}{2015}).

\bibitem{Seward1984}
\bibinfo{author}{{Seward}, F.~D. and {Harnden}, Jr., F.~R. and {Helfand}, D.~J.}
\newblock \bibinfo{title}{Discovery of a 50 millisecond pulsar in the Large Magellanic Cloud}
\newblock \emph{\bibinfo{journal}{Astrophys. J. Lett.}}
  \textbf{\bibinfo{volume}{287}}, \bibinfo{pages}{L19-L22}
  (\bibinfo{year}{1984}).

\bibitem{Gotthelf2000}
\bibinfo{author}{{Gotthelf}, E.~V. and {Wang}, Q.~D.}
\newblock \bibinfo{title}{A Spatially Resolved Plerionic X-Ray Nebula around PSR B0540-69}
\newblock \emph{\bibinfo{journal}{Astrophys. J. Lett.}}
\textbf{\bibinfo{volume}{532}}, \bibinfo{pages}{L117-L120}
(\bibinfo{year}{2000}).

\bibitem{Petre2007}
\bibinfo{author}{{Petre}, R. and {Hwang}, U. and {Holt}, S.~S. and {Safi-Harb}, S. and {Williams}, R.~M.}
\newblock \bibinfo{title}{The X-Ray Structure and Spectrum of the Pulsar Wind Nebula Surrounding PSR B0540-69.3}
\newblock \emph{\bibinfo{journal}{Astrophys. J.}}
\textbf{\bibinfo{volume}{662}}, \bibinfo{pages}{988-997}
(\bibinfo{year}{2007}).

\bibitem{Zhang2001}
\bibinfo{author}{{Zhang}, W. and {Marshall}, F.~E. and {Gotthelf}, E.~V. and {Middleditch}, J. and {Wang}, Q.~D.}
\newblock \bibinfo{title}{A Phase-connected Braking Index Measurement for the Large Magellanic Cloud Pulsar PSR B0540-69}
\newblock \emph{\bibinfo{journal}{Astrophys. J. Lett.}}
\textbf{\bibinfo{volume}{554}}, \bibinfo{pages}{L177-L180}
(\bibinfo{year}{2001}).

\bibitem{Ge2012}
\bibinfo{author}{{Ge}, M.~Y. and {Lu}, F.~J. and {Qu}, J.~L. and {Zheng}, S.~J. and {Chen}, Y. and {Han}, D.~W.}
\newblock \bibinfo{title}{X-Ray Phase-resolved Spectroscopy of PSRs B0531+21, B1509-58, and B0540-69 with RXTE}
\newblock \emph{\bibinfo{journal}{Astrophys. J. Suppl.}}
  \textbf{\bibinfo{volume}{199}}, \bibinfo{pages}{32}
  (\bibinfo{year}{2012}).

\bibitem{Ferdman2015}
\bibinfo{author}{{Ferdman}, R.~D. and {Archibald}, R.~F. and {Kaspi}, V.~M.}
\newblock \bibinfo{title}{Long-term Timing and Emission Behavior of the Young Crab-like Pulsar PSR B0540-69}.
\newblock \emph{\bibinfo{journal}{Astrophys. J. Lett.}}
  \textbf{\bibinfo{volume}{812}}, \bibinfo{pages}{L95}
  (\bibinfo{year}{2015}).

\bibitem{Marshall2016}
\bibinfo{author}{{Marshall}, F.~E. and {Guillemot}, L. and {Harding}, A.~K. and {Martin}, P. and {Smith}, D.~A.}
\newblock \bibinfo{title}{A New, Low Braking Index for the LMC Pulsar B0540-69}.
\newblock \emph{\bibinfo{journal}{Astrophys. J. Lett.}}
  \textbf{\bibinfo{volume}{827}}, \bibinfo{pages}{L39}
  (\bibinfo{year}{2016}).

\bibitem{Alpar1984a}
\bibinfo{author} {{Alpar} M.~A., {Anderson} P. W., {Shaham} J.}
\newblock \bibinfo{title}{Vortex creep and the internal temperature of neutron stars. I - General theory}.
\newblock \emph{\bibinfo{journal}{Astrophys. J.}}
 \textbf{\bibinfo{volume}{276}}, \bibinfo{pages}{325-334}
  (\bibinfo{year}{1984}).

\bibitem{Espinoza2011}
\bibinfo{author}{{Espinoza}, C.~M. and {Lyne}, A.~G. and {Stappers}, B.~W. and {Kramer}, M.}
\newblock \bibinfo{title}{{A study of 315 glitches in the rotation of 102 pulsars.}}
\newblock \emph{\bibinfo{journal}{Mon. Not. R. Astron. Soc.}}
  \textbf{\bibinfo{volume}{414}}, \bibinfo{pages}{1679-1704}
  (\bibinfo{year}{2011}).

\bibitem{Kargaltsev15}
\bibinfo{author}{Kargaltsev, O., Cerutti, B., Lyubarsky, Y., \& Striani, E.}
\newblock \bibinfo{title}{Pulsar-Wind Nebulae. Recent Progress in Observations and Theory}.
\newblock \emph{\bibinfo{journal}{Space Sci. Rev.}}
  \textbf{\bibinfo{volume}{191}}, \bibinfo{pages}{391-439}
  (\bibinfo{year}{2015}).


\bibitem{Bucciantini11}
\bibinfo{author}{Bucciantini, N., Arons, J., \& Amato, E.}
\newblock \bibinfo{title}{Modelling spectral evolution of pulsar wind nebulae inside supernova remnants}.
\newblock \emph{\bibinfo{journal}{Mon. Not. R. Astron. Soc.}}
  \textbf{\bibinfo{volume}{410}}, \bibinfo{pages}{381-398}
  (\bibinfo{year}{2011}).

\bibitem{Wang16a}
\bibinfo{author}{Wang, L. J., Dai, Z. G., Liu, L. D., \& Wu, X. F.}
\newblock \bibinfo{title}{Probing the Birth of Post-merger Millisecond Magnetars with X-Ray and Gamma-Ray Emission}.
\newblock \emph{\bibinfo{journal}{Astrophys. J.}}
  \textbf{\bibinfo{volume}{823}}, \bibinfo{pages}{15}
  (\bibinfo{year}{2016}).

\bibitem{Tavani2011}
\bibinfo{author}{{Tavani}, M. and {Bulgarelli}, A. and {Vittorini}, V. and {Pellizzoni}, A. and
        {Striani}, E. et al. }
\newblock \bibinfo{title}{Discovery of Powerful Gamma-Ray Flares from the Crab Nebula}.
\newblock \emph{\bibinfo{journal}{Science}}
  \textbf{\bibinfo{volume}{331}}, \bibinfo{pages}{736-739}
  (\bibinfo{year}{2011}).

\bibitem{Abdo2011}
\bibinfo{author}{{Abdo}, A.~A. and {Ackermann}, M. and {Ajello}, M. and {Allafort}, A. and {Baldini}, L. et al. }
\newblock \bibinfo{title}{Gamma-Ray Flares from the Crab Nebula}.
\newblock \emph{\bibinfo{journal}{Science}}
  \textbf{\bibinfo{volume}{331}}, \bibinfo{pages}{739-742}
  (\bibinfo{year}{2011}).

\bibitem{Reynolds2018}
\bibinfo{author}{{Reynolds}, S.~P. and {Borkowski}, K.~J. and {Gwynne}, P.~H.}
\newblock \bibinfo{title}{Expansion and Brightness Changes in the Pulsar-wind Nebula in the Composite Supernova Remnant Kes 75}.
\newblock \emph{\bibinfo{journal}{Astrophys. J.}}
  \textbf{\bibinfo{volume}{856}}, \bibinfo{pages}{133}
  (\bibinfo{year}{2018}).

\bibitem{Pavlov2003}
\bibinfo{author}{{Pavlov}, G.~G. and {Teter}, M.~A. and {Kargaltsev}, O. and {Sanwal}, D.}
\newblock \bibinfo{title}{The Variable Jet of the Vela Pulsar}.
\newblock \emph{\bibinfo{journal}{Astrophys. J.}}
  \textbf{\bibinfo{volume}{591}}, \bibinfo{pages}{1157-1171}
  (\bibinfo{year}{2003}).

\bibitem{Lyne2010}
\bibinfo{author}{{Lyne}, A. and {Hobbs}, G. and {Kramer}, M. and {Stairs}, I. and {Stappers}, B.}
\newblock \bibinfo{title}{Switched Magnetospheric Regulation of Pulsar Spin-Down}.
\newblock \emph{\bibinfo{journal}{Science}}
  \textbf{\bibinfo{volume}{329}}, \bibinfo{pages}{408-412}
  (\bibinfo{year}{2010}).

\bibitem{Kramer2006}
\bibinfo{author}{{Kramer}, M. and {Lyne}, A. G. and {O'Brien}, J. T. and {Jordan}, C. A.  and {Lorimer}, D. R.}
\newblock \bibinfo{title}{A periodically-active pulsar giving insight into magnetospheric physics}.
\newblock \emph{\bibinfo{journal}{Science}}
  \textbf{\bibinfo{volume}{314}}, \bibinfo{pages}{97}
  (\bibinfo{year}{2006}).

\bibitem{Allafort13}
\bibinfo{author}{{Allafort}, A. et al. }
\newblock \bibinfo{title}{PSR J2021+4026 in the Gamma Cygni Region: The First Variable $\gamma$-Ray Pulsar Seen by the Fermi LAT}.
\newblock \emph{\bibinfo{journal}{Astrophys. J. Lett.}}
  \textbf{\bibinfo{volume}{777}}, \bibinfo{pages}{L2}
  (\bibinfo{year}{2013}).

\bibitem{Goldreich1969}
\bibinfo{author}{{Goldreich}, P. and {Julian}, W.~H.}
\newblock \bibinfo{title}{Pulsar Electrodynamics}
\newblock \emph{\bibinfo{journal}{Astrophys. J.}}
  \textbf{\bibinfo{volume}{157}}, \bibinfo{pages}{869-880}
  (\bibinfo{year}{1969}).

\bibitem{Ruderman1975}
\bibinfo{author}{{Ruderman}, M.~A. and {Sutherland}, P.~G.}
\newblock \bibinfo{title}{Theory of pulsars - Polar caps, sparks, and coherent microwave radiation}
\newblock \emph{\bibinfo{journal}{Astrophys. J.}}
  \textbf{\bibinfo{volume}{196}}, \bibinfo{pages}{51-72}
  (\bibinfo{year}{1975}).

\bibitem{Arons1979}
\bibinfo{author}{{Arons}, J. and {Scharlemann}, E.~T.}
\newblock \bibinfo{title}{Pair formation above pulsar polar caps - Structure of the low altitude acceleration zone}
\newblock \emph{\bibinfo{journal}{Astrophys. J.}}
  \textbf{\bibinfo{volume}{231}}, \bibinfo{pages}{854-879}
  (\bibinfo{year}{1979}).

\bibitem{Cheng1986}
\bibinfo{author}{{Cheng}, K.~S. and {Ho}, C. and {Ruderman}, M.}
\newblock \bibinfo{title}{Energetic radiation from rapidly spinning pulsars. I - Outer magnetosphere gaps. II - VELA and Crab}
\newblock \emph{\bibinfo{journal}{Astrophys. J.}}
  \textbf{\bibinfo{volume}{300}}, \bibinfo{pages}{500-539}
  (\bibinfo{year}{1986}).

\bibitem{Kou2016}
\bibinfo{author}{{Kou}, F. F. and {Ou}, Z. W. and {Tong}, H.}
\newblock \bibinfo{title}{On the variable timing behavior of PSR B0540-69: an almost excellent example to study the pulsar braking mechanism.}
\newblock \emph{\bibinfo{journal}{RAA}}
\textbf{\bibinfo{volume}{16}}, \bibinfo{pages}{79}
(\bibinfo{year}{2016}).

\bibitem{Eksi17}
\bibinfo{author}{Ek\c{s}i, K. Y.}
\newblock \bibinfo{title}{On the new braking index of PSR B0540-69: further support for magnetic field growth of neutron stars following submergence by fallback accretion}.
\newblock \emph{\bibinfo{journal}{Mon. Not. R. Astron. Soc.}}
  \textbf{\bibinfo{volume}{469}}, \bibinfo{pages}{1974-1978}
  (\bibinfo{year}{2017}).

\bibitem{Bucciantini06}
\bibinfo{author}{Bucciantini, N., Thompson, T. A., Arons, J., Quataert, E., Del Zanna, L.}
\newblock \bibinfo{title}{Relativistic magnetohydrodynamics winds from rotating neutron stars}.
\newblock \emph{\bibinfo{journal}{Mon. Not. R. Astron. Soc.}}
  \textbf{\bibinfo{volume}{368}}, \bibinfo{pages}{1717-1734}
  (\bibinfo{year}{2006}).




\bibitem{Mignani2010}
\bibinfo{author}{{Mignani}, R.~P. and {Sartori}, A. and {de Luca}, A. and {Rudak}, B. and {S{\l}owikowska}, A. and {Kanbach}, G. and {Caraveo}, P.~A.}
\newblock \bibinfo{title}{HST/WFPC2 observations of the LMC pulsar PSR B0540-69}.
\newblock \emph{\bibinfo{journal}{Astron. Astrophys.}}
  \textbf{\bibinfo{volume}{515}}, \bibinfo{pages}{A110}
  (\bibinfo{year}{2010}).

\bibitem{Jahoda2006}
\bibinfo{author}{{Jahoda}, K. and {Markwardt}, C.~B. and {Radeva}, Y. and {Rots}, A.~H. and {Stark}, M.~J. and {Swank}, J.~H. and {Strohmayer}, T.~E. and {Zhang}, W.}
\newblock \bibinfo{title}{Calibration of the Rossi X-Ray Timing Explorer Proportional Counter Array}.
\newblock \emph{\bibinfo{journal}{Astrophys. J. Suppl.}}
  \textbf{\bibinfo{volume}{163}}, \bibinfo{pages}{401-423}
  (\bibinfo{year}{2006}).

\bibitem{Jansen2001}
\bibinfo{author}{{Jansen}, F. and {Lumb}, D. and {Altieri}, B. and {Clavel}, J. and {Ehle}, M. and {Erd}, C. and {Gabriel}, C. and {Guainazzi}, M. and {Gondoin}, P. and {Much}, R. and {Munoz}, R. and {Santos}, M. and {Schartel}, N. and {Texier}, D. and {Vacanti}, G.}
\newblock \bibinfo{title}{XMM-Newton observatory. I. The spacecraft and operations}.
\newblock \emph{\bibinfo{journal}{Astron. Astrophys.}}
  \textbf{\bibinfo{volume}{365}}, \bibinfo{pages}{1-6}
  (\bibinfo{year}{2001}).

\bibitem{Harrison2013}
\bibinfo{author}{{Harrison}, F.~A. and {Craig}, W.~W. and {Christensen}, F.~E. and
	{Hailey}, C.~J. and {Zhang}, W.~W., et al.}
\newblock \bibinfo{title}{The Nuclear Spectroscopic Telescope Array (NuSTAR) High-energy X-Ray Mission}.
\newblock \emph{\bibinfo{journal}{Astrophys. J.}}
  \textbf{\bibinfo{volume}{770}}, \bibinfo{pages}{103}
  (\bibinfo{year}{2013}).

\bibitem{Gehrels2004}
\bibinfo{author}{{Gehrels}, N. and {Chincarini}, G. and {Giommi}, P. and {Mason}, K.~O. and 	 {Nousek}, J.~A. et al.}
\newblock \bibinfo{title}{The Swift Gamma-Ray Burst Mission}.
\newblock \emph{\bibinfo{journal}{Astrophys. J.}}
  \textbf{\bibinfo{volume}{611}}, \bibinfo{pages}{1005-1020}
  (\bibinfo{year}{2004}).

\bibitem{Campana2008}
\bibinfo{author}{{Campana}, R. and {Mineo}, T. and {de Rosa}, A. and {Massaro}, E. and {Dean}, A.~J. and {Bassani}, L.}
\newblock \bibinfo{title}{X-ray observations of the Large Magellanic Cloud pulsar PSR B0540-69 and its pulsar wind nebula}.
\newblock \emph{\bibinfo{journal}{Mon. Not. R. Astron. Soc.}}
  \textbf{\bibinfo{volume}{389}}, \bibinfo{pages}{691-700}
  (\bibinfo{year}{2008}).

\bibitem{TEMPO2}
\bibinfo{author}{{Hobbs}, G.~B.}, \bibinfo{author}{{Edwards}, R.~T.} \&
  \bibinfo{author}{{Manchester}, R.~N.}
\newblock \bibinfo{title}{{TEMPO2, a new pulsar-timing package - I. An
  overview}}.
\newblock \emph{\bibinfo{journal}{Mon. Not. R. Astron. Soc.}}
  \textbf{\bibinfo{volume}{369}}, \bibinfo{pages}{655--672}
  (\bibinfo{year}{2006}).

\bibitem{Mignani2018}
\bibinfo{author}{{Mignani}, R. P. and {Shearer}, A. and {de Luca}, A. and {Marshall}, F. E. and {Guillemot}, L. and {Smith},  D. A. and {Rudak}, B. and {Zampieri}, L. and {Barbieri}, C. and {Naletto}, G. and {Gouiffes}, C. and {Kanbach}, G.}
\newblock \bibinfo{title}{The first ultraviolet detection of the Large Magellanic Cloud pulsar PSR B0540-69}.
\newblock \emph{\bibinfo{journal}{Astrophys. J.}}
  \textbf{\bibinfo{volume}{871}}, \bibinfo{pages}{246}
  (\bibinfo{year}{2019}).

\end{thebibliography}
\end{document}